\documentclass[pre,floatfix,reprint,superscriptaddress,amsmath,amssymb]{revtex4-2}

\usepackage[T1]{fontenc}
\usepackage{amsmath,amssymb,bm,mathtools}
\usepackage{booktabs,array,multirow}
\usepackage{graphicx}
\usepackage{placeins}
\usepackage{microtype}
\usepackage{siunitx}
\usepackage{enumitem}
\usepackage{xcolor}
\usepackage[normalem]{ulem}
\usepackage{xurl}
\usepackage[colorlinks=true,linkcolor=blue,citecolor=blue,urlcolor=blue]{hyperref}

\graphicspath{{./}}
\microtypesetup{protrusion=true,expansion=true}
\newcommand{\avg}[1]{\left\langle #1\right\rangle}
\newcommand{\dd}{\mathrm{d}}
\newcommand{\ii}{\mathrm{i}}
\newcommand{\kB}{k_{\mathrm B}}
\newcommand{\vect}[1]{\bm{#1}}
\newcommand{\Var}{\operatorname{Var}}

\newcommand{\erf}{\operatorname{erf}}
\newcommand{\order}{\mathcal{O}}

\begin{document}

\title{Collective Ion Dynamics from Finite-Volume Fluctuations in Model Explicit-Solvent Electrolytes}

\author{Jeongmin Kim}
\email{jeongmin@pusan.ac.kr}
\affiliation{Department of Chemistry Education, Graduate Department of Chemical
Materials, and Institute for Plastic Information and Energy Materials,
Pusan National University, Busan 46241, Republic of Korea}

\begin{abstract}
Understanding how collective ion transport emerges from equilibrium fluctuations is central to electrolyte statistical mechanics. Finite-volume fluctuations provide an accessible route to this information, but their interpretation is complicated because they mix wave numbers and collective fields. Here, we extend the finite-volume counting framework for inferring collective ion diffusion by combining the exact window projection with an inertial independent-particle reference, and apply it to a symmetric 1:1 solvent primitive model. The growth of ion-number fluctuations between the ballistic and plateau regimes appears nearly ideal, but this apparent ideality results from compensation between negative structural and positive dynamical excess contributions. The coupled ion-number--solvent relaxation further reveals a signed redistribution between solvent-associated and solvent-orthogonal projections that is largely hidden in the total response. The inferred collective diffusion depends on observation length and dynamical closure: measured structure alone does not systematically improve the estimate, whereas the coupled-field description reduces the high-concentration discrepancy. These results provide a basis for analyzing more realistic electrolytes with mutually coupled ion-number, charge, and solvent fluctuations.
\end{abstract}


\maketitle

\section{Introduction}
Collective dynamics govern how concentration and charge fluctuations decay toward equilibrium, closely related to mass transport and electrostatic screening in electrolytes~\cite{Onsager1931I,Green1954,Kubo1957,HansenMcDonald}. In the dilute limit, the solvent can often be treated as a homogeneous background medium, as in Debye--H\"uckel theory for electrostatic screening~\cite{DebyeHuckel1923}. As concentration increases, however, correlations among ions and solvent particles become increasingly important for collective relaxation~\cite{HansenMcDonald,Fong2020Transport,Roling2024Dynamics}, linking microscopic fluctuations to macroscopic transport~\cite{Fong2020Transport,Balos2020Conductivity}. Understanding this connection is necessary for interpreting collective transport in electrolytes, where ion-number and charge-density fluctuations represent distinct collective variables with different relaxation mechanisms.

The conventional approach based on finite-wave-number and frequency-dependent response provides the natural framework for describing collective relaxation in liquids~\cite{Caillol1987FiniteWave,Minh2023ElectricalNoise,Minh2026Asymmetric}. The static structure factor determines the equilibrium amplitude of each density fluctuation, whereas the intermediate scattering function, or equivalently its frequency-domain representation, the dynamic structure factor, describes its relaxation~\cite{vanHove1954,HansenMcDonald}. Their dependence on wave number resolves the relevant length scales and provides information about collective diffusion, structural narrowing, coupling among conserved variables, and dynamical memory~\cite{deGennes1959,Mori1965}. This reciprocal-space description therefore provides a detailed connection between equilibrium liquid structure and collective transport.

As a complementary real-space approach, finite-volume counting has recently attracted renewed attention for probing collective dynamics from particle-number fluctuations within a prescribed observation volume~\cite{SvedbergInouye1911,Smoluchowski1916Counting,MartinYalcin1980,Lebowitz1983Charge,KimLuijtenFisher2005,KimFisher2008Charge,MinhRotenbergMarbach2023,Mackay2024,Carter2025}. Such fluctuations are directly accessible in particle-resolved experiments and simulations, and their static variance is connected to finite-volume thermodynamics and Kirkwood--Buff theory~\cite{KirkwoodBuff1951,Kruger2013,Dawass2018,SevillaCortes2022,Simon2022}. Their time-dependent mean-squared increment, as formalized in the recent Countoscope framework, provides information about collective transport without requiring persistent tracking of the particles inside the observation volume~\cite{Mackay2024,Carter2025}. A recent intensity-based extension further extracts particle dynamics from microscopy images without resolving individual particles~\cite{Hermann2026IntensityCountoscope}. This advantage comes with an inherent interpretative challenge: a finite observation window projects a range of wave numbers onto a single time-dependent observable, particularly in strongly interacting systems such as concentrated electrolytes~\cite{MinhRotenbergMarbach2023}. Its interpretation therefore requires a separation between the exact geometrical projection of the measured fluctuations and the structural and dynamical assumptions used to infer transport.

The structural and dynamical models required to interpret finite-volume relaxation are closely related to the closures introduced in stochastic density-functional theories (SDFTs)~\cite{Dean1996,Illien2025,DemeryDean2016,BernardJardatRotenbergIllien2023,Bonneau2024Frequency,BonneauDemeryRaphael2025}. Recent explicit polar-solvent extensions describe ions and solvent dipoles and resolve coupled charge and polarization relaxation~\cite{IllienCarofRotenberg2024}. The importance of the solvent field is highlighted by the memory-dependent ionic response obtained upon eliminating solvent polarization, with two-step relaxation predicted for slowly relaxing solvents~\cite{VargheseRotenbergIllien2026}. Solvent dielectric response and spatial correlations also influence ionic screening and nonlinear conductivity~\cite{Kornyshev1983Nonlocal,Berthoumieux2024Nonlinear,DemeryToquer2026Solvent}.

Here, we formulate and test an explicit-solvent extension of finite-volume fluctuation analysis using molecular dynamics of a symmetric 1:1 solvent primitive model. Cation--anion exchange symmetry in the studied model provides a controlled setting in which ion-number fluctuations can couple to solvent-density fluctuations, while charge cross correlations with both densities vanish~\cite{BhatiaThornton1970,CiachGozdzStell2007}. To describe the molecular-dynamics response, we extend the independent-particle reference to underdamped dynamics and remove the leading volume-fluctuation contribution associated with \(\mathrm{NpT}\) sampling. By separating the exact finite-window projection from the structural and dynamical closures used for transport inference, we show that the apparently ideal ion-number response in the intermediate relaxation regime between ballistic growth and the long-time plateau results from compensation between structural and dynamical nonideal contributions. The coupled ion-number--solvent relaxation further reveals a signed redistribution between solvent-associated and solvent-orthogonal projections, whose intermediate-concentration turnover is partly captured by an ideal-composition Bhatia--Thornton reference. The resulting scale-dependent collective diffusion estimates demonstrate the importance of choosing appropriate structural and dynamical closures when inferring collective transport from finite-volume observations.

This paper is organized as follows. Section II presents the finite-volume fluctuation theory and its extension to underdamped molecular dynamics. Section III describes the simulation model and computational methods, including the \(\mathrm{NpT}\) volume-fluctuation correction. Section IV presents the simulation results and analyzes the time-dependent finite-volume fluctuations and collective diffusion estimates. Section V summarizes the main conclusions.

\section{Finite-volume counting theory}\label{sec:theory}

We first state the exact relation between finite-volume counts and collective
correlations, then introduce a minimal dynamical closure. The former is a
measurement identity; only the latter assumes a propagator and can therefore
carry constitutive error.

\subsection{finite-volume fluctuations and their relation to wavevector-resolved correlations}
\label{sec:theory-observables}

The ion-number and valence-charge counts within an observation region \(\Lambda\)\textcolor{black}{~\cite{Mackay2024,Carter2025,MinhRotenbergMarbach2023}} are defined as
\begin{equation}
X_\Lambda(t)=\int\dd^3r\,w_\Lambda(\vect r)\rho_X(\vect r,t),
\label{eq:microscopic_fields}
\end{equation}
where \(w_\Lambda(\vect r)\) is a window function equal to 1 when \(\vect r\) lies inside \(\Lambda\) and zero otherwise. The microscopic field is \(\rho_X(\vect r,t)=\sum_{i\in\mathrm{ions}}a_i^X\delta[\vect r-\vect r_i(t)]\), with \(a_i^N=1\), \(a_i^Z=z_i=\pm1\), and \(X=N,Z\). Analogously, the solvent count \(S_\Lambda(t)\) is defined using \(\rho_S(\vect r,t)=\sum_{j\in\mathrm{solv}}\delta[\vect r-\vect r_j(t)]\).

The mean-squared fluctuation (MSF) of finite-volume variable in thermal equilibrium obeys the following exact relation: 
\begin{equation}
 \begin{aligned}
 M_X(t)&\equiv\avg{[\delta X_\Lambda(t)-\delta X_\Lambda(0)]^2}\\
 &=2[C_X(0)-C_X(t)]\\
 &=2\int\frac{\dd\vect k\,\dd\vect q}{(2\pi)^6}
 \widetilde w_\Lambda(-\vect k)\widetilde w_\Lambda(-\vect q)\\[-0.25em]
 &\qquad\times\Bigl[
 \avg{\delta\rho_X(\vect k,0)\delta\rho_X(\vect q,0)}\\[-0.25em]
 &\qquad\phantom{\times\Bigl[}{}-\avg{\delta\rho_X(\vect k,t)\delta\rho_X(\vect q,0)}\Bigr]\\
 &=2\rho_{\mathrm{ion}}\int\frac{\dd\vect k}{(2\pi)^3}
 f_\Lambda(\vect k)[F_{XX}(k,0)-F_{XX}(k,t)]\\
 &=2\rho_{\mathrm{ion}}\int\frac{\dd\vect k}{(2\pi)^3}
 f_\Lambda(\vect k)S_{XX}(k)[1-\Phi_X(k,t)]\\
 &=2\rho_{\mathrm{ion}}
 \avg{S_{XX}(k)[1-\Phi_X(k,t)]}_\Lambda
 \end{aligned}
 \label{eq:exact_msf_bulk}
\end{equation}
Here, \(\delta X_\Lambda(t)=X_\Lambda-\avg{X_\Lambda}=\int\dd\vect k\,
\widetilde w_\Lambda(-\vect k)\delta\rho_X(\vect k,t)/(2\pi)^3\),
\(C_X(t)\equiv\avg{\delta X_\Lambda(t)\delta X_\Lambda(0)}\), \(S_{XX}(k)\equiv F_{XX}(k,0)\), and \(\Phi_X(k,t)\equiv F_{XX}(k,t)/S_{XX}(k)\). \(\avg{\cdots}\) denotes the ensemble average.  Translational invariance then gives \(\avg{\delta\rho_X(\vect k,t)\delta\rho_X(\vect q,0)}=(2\pi)^3\rho_{\mathrm{ion}}\delta(\vect k+\vect q)F_{XX}(k,t)\), which enforces \(\vect q=-\vect k\). The last equality defines the compact finite-volume average \(\avg{\cdots}_\Lambda\):
\begin{equation}
 \begin{aligned}
 \avg{A}_\Lambda&\equiv
 \int\frac{\dd\vect k}{(2\pi)^3}f_\Lambda(\vect k)A(\vect k)
 \end{aligned}
 \label{eq:window_average}
\end{equation}
with $f_\Lambda(\vect k)=|\widetilde w_\Lambda(\vect k)|^2$, and $\widetilde w_\Lambda(\vect k)=
 \int\dd^3r\,w_\Lambda(\vect r)e^{-\ii\vect k\cdot\vect r}$.

For the symmetric solvent primitive model studied here, \(N_\Lambda=N_{+,\Lambda}+N_{-,\Lambda}\) and \(Z_\Lambda=N_{+,\Lambda}-N_{-,\Lambda}\) define the ion-number and valence-charge variables, respectively. Cation--anion exchange symmetry makes \(N\) and \(S\) even and \(Z\) odd. The ion-number and solvent fluctuations can therefore be analyzed as a coupled \(N\)--\(S\) block, separately from the charge sector. In this work, we focus on the finite-volume counting of ion number for $X=N$, and its coupling to solvent fluctuations.

\subsection{Minimal dynamical model for molecular dynamics}
\label{sec:theory-linear}

The MSF \(M_X(t)\) evolves from zero toward its long-time plateau \(2C_X(0)=2\Var(X_\Lambda)\), which is related to finite-volume Kirkwood--Buff analysis~\cite{Kruger2013}. According to Eq.~\eqref{eq:exact_msf_bulk}, finite-volume counting is directly connected to wave-vector-resolved correlation functions~\cite{vanHove1954,HansenMcDonald}. The functional \(M_X(t)\equiv\mathcal M_X[S_{XX},\Phi_X]\) separates the exact finite-volume projection from the chosen static and dynamical inputs. Motivated by de Gennes narrowing and dynamical density-functional descriptions of density relaxation~\cite{deGennes1959,MarconiTarazona1999,ArcherEvans2004}, we assume a single-rate decay of \(\Phi_N\):
\begin{equation}
\Phi_N^{\mathrm{UD}}(k,t)\simeq e^{-\Gamma_N(k)g(t;\tau_v)}.
\label{eq:inertial_phi}
\end{equation}
Here, \(\Gamma_N(k)=D_N^Jk^2/S_{NN}(k)\), \(g(t;\tau_v)=t-\tau_v(1-e^{-t/\tau_v})\), and \(\tau_v=mD_N^J/(\kB T)\). The independently measured collective diffusion coefficient \(D_N^J\) sets both the decay rate and the closure-imposed velocity-relaxation time \(\tau_v\). The superscript \(\mathrm{UD}\) denotes the underdamped clock used to include short-time inertial motion in molecular dynamics. The clock reproduces the equal-mass short-time second moment but is not an independently measured collective velocity-relaxation time~\cite{UhlenbeckOrnstein1930}. Molecular friction and generalized Langevin descriptions can instead involve a time-dependent memory kernel; the single clock used here is a minimal closure rather than a general representation of that memory~\cite{Straube2020Friction,Ayaz2022Memory}. For overdamped Brownian dynamics, \(g(t)=t\), and  $\Phi_N$ reduces to the expression used in the previous work~\cite{Mackay2024,Carter2025}.

For a cubic observation volume of a side length $L_{\mathrm{obs}}$, the analytical ideal-number MSF is obtained by inserting \(S_{NN}(k)=1\) into Eq.~\eqref{eq:exact_msf_bulk}~\cite{Mackay2024,Carter2025}:
\begin{equation}
\frac{M_N^{\mathrm{id}}(t)}{2\avg{N_\Lambda}}=1-h^3(u),
\label{eq:number_ideal_analytic}
\end{equation}
where $h(u)=\erf(u^{-1/2})+\sqrt{\frac{u}{\pi}}\left(e^{-1/u}-1\right)$, and $u=4D_N^Jg(t;\tau_v)/L_{\mathrm{obs}}^2$. This analytical expression provides a means of estimating \(D_N^J\) by fitting \(M_N^{\mathrm{id}}(t;D)\) to the computed \(M_N(t)\).

The leading-order ballistic growth of \(M_N^{\mathrm{id}}(t)\), however, is common across the studied conditions despite their different structural and collective dynamical correlations. The short-time expansion of Eq.~\eqref{eq:number_ideal_analytic} is
\begin{equation}
\frac{M_N^{\mathrm{id}}(t)}
{2\langle N_\Lambda\rangle}=
\frac{6t}{L_{\mathrm{obs}}}
\sqrt{\frac{\kB T}{2\pi m}}
+\order(t^2),
\label{eq:sharp-window-ballistic}
\end{equation}
where \(h(u)=1-\sqrt{u/\pi}+\order(u)\) and \(g(t;\tau_v)=t^2/(2\tau_v)+\order(t^3)\). Here, \(\langle N_\Lambda\rangle=\rho_{\mathrm{ion}}L_{\mathrm{obs}}^3=2\rho_{\mathrm{salt}}L_{\mathrm{obs}}^3\) for the symmetric 1:1 electrolyte. Thus, the initial ballistic growth is proportional to \(t\) and exhibits surface-area scaling, \(M_N^{\mathrm{id}}(t)\sim L_{\mathrm{obs}}^2\). Equation~\eqref{eq:sharp-window-ballistic} is independent of \(D_N^J\) and applies to the leading-order behavior of the computed \(M_N(t)\) as well. This is because the high-wave-number limit \(S_{NN}(k)\to1\) makes the leading coefficient insensitive to static correlations, while \(\tau_v=mD_N^J/(\kB T)\) causes \(D_N^J\) to cancel from it.

\subsection{Decomposition of nonideal contributions to \(M_N(t)\) and ion-number--solvent coupling}
\label{sec:theory-number-decomposition}

The total excess, \(M_N^{\mathrm{ex}}(t)\equiv M_N(t)-M_N^{\mathrm{id}}(t)\), is decomposed as
\begin{equation}
M_N(t)=M_N^{\mathrm{id}}(t)+\Delta M_N^{\mathrm{str}}(t)
+\Delta M_N^{\mathrm{dyn}}(t)+\Delta M_N^{\mathrm{res}}(t).
\label{eq:number_excess_all_components}
\end{equation}
This decomposition separates the contributions associated with static structure and dynamical correlations. The residual quantifies incomplete reciprocal-space reconstruction rather than an additional physical mechanism; in the numerical analysis, it accounts for the contribution not represented by the sampled wavevectors. Using the functional \(\mathcal M_N[S_{NN},\Phi_N]\), the individual terms are defined by
\begin{subequations}
\label{eq:number_excess_component_definitions}
\begin{align}
M_N(t)
&=\mathcal M_N[1,\Phi^0_N] \\
&\quad+\left(\mathcal M_N[S_{NN},\Phi_N^{\mathrm{UD}}]
-\mathcal M_N[1,\Phi^0_N]\right) \\
&\quad+\left(\mathcal M_N[S_{NN},\Phi_N^{\mathcal K}]
-\mathcal M_N[S_{NN},\Phi_N^{\mathrm{UD}}]\right) \\
&\quad+\left(M_N(t)-\mathcal M_N[S_{NN},\Phi_N^{\mathcal K}]\right),
\end{align}
\end{subequations}
where the four terms are \(M_N^{\mathrm{id}}\), \(\Delta M_N^{\mathrm{str}}\), \(\Delta M_N^{\mathrm{dyn}}\), and \(\Delta M_N^{\mathrm{res}}\), respectively. Here, \(\mathcal K\) denotes the set of wave vectors used to calculate the reciprocal-space correlation functions, and the independent-particle propagator is \(\Phi^0_N(k,t)=\exp[-D_N^Jk^2g(t;\tau_v)]\). As is clear in Eq.~\eqref{eq:number_excess_component_definitions}, \(\Delta M_N^{\mathrm{str}}\) is the complete change produced by introducing the measured \(S_{NN}(k)\) into both the static amplitude and the structure-informed \(k^2/S_{NN}(k)\) relaxation rate, whereas \(\Delta M_N^{\mathrm{dyn}}\) measures the subsequent departure from this single-rate response. For the symmetric 1:1 solvent electrolytes studied here, \(S_{NN}(k)\) approaches its ideal high-wave-number limit, \(S_{NN}(k)\to1\), whereas its low-wave-number limit is close to the ideal-composition/Carnahan--Starling reference~\cite{BhatiaThornton1970,CarnahanStarling1969} (see Sec.~S2 in the SM~\cite{SupplementalMaterial}). These limiting behaviors characterize the static ion-number structure but do not resolve how ion-number fluctuations couple to solvent-density fluctuations during relaxation. We therefore consider the full \(N\)--\(S\) structure and relaxation matrices, whose off-diagonal elements \(S_{NS}(k)\) and \(F_{NS}(k,t)\) quantify the static and dynamical ion--solvent correlations, respectively (see Sec.~S6 in the SM~\cite{SupplementalMaterial}).

To resolve this coupling, we decompose the ion-number fluctuation field into its static projection onto the solvent-density field, denoted solvent-associated (\(\parallel\)), and the covariance-orthogonal remainder, denoted solvent-orthogonal (\(\perp\)): \(\delta\rho_N=\delta\rho_N^\parallel+\delta\rho_N^\perp\). Because the two fields are covariance-orthogonal at equal time, their static contributions satisfy
\begin{equation}
\begin{aligned}
S_{NN}(k)
&=S_{NN}^{\parallel}(k)+S_{NN}^{\perp}(k)\\
&=w_\parallel(k)S_{NN}(k)+w_\perp(k)S_{NN}(k).
\end{aligned}
\end{equation}
where 
\begin{equation}
 w_\parallel(k)=|c_{NS}(k)|^2=\frac{S^2_{NS}(k)}
 {S_{NN}(k)S_{SS}(k)}, \label{eq:even_static_matrix}
\end{equation}
and $w_\perp(k)=1-w_\parallel(k)$. For the dynamical correlations, we define \(F_\perp(k,t)\equiv F_{NN}^{\perp\perp}(k,t)\) and \(F_\parallel(k,t)\equiv F_{NN}^{\parallel}(k,t)=F_{NN}(k,t)-F_{NN}^{\perp\perp}(k,t)\). Rather than treating these projected responses as exact dynamical normal modes, we use them to decompose the dynamical nonideal contribution:
\begin{equation}
\Delta M_N^{\mathrm{dyn}}(t)
=\sum_{\alpha=\perp,\parallel}\Delta M_N^{\mathrm{dyn},\alpha}(t)
\end{equation}
\begin{equation}
\Delta M_N^{\mathrm{dyn},\alpha}(t)
=2\rho_{\mathrm{ion}}
\left\langle F_\alpha^{\mathrm{UD}}(k,t)-F_\alpha(k,t)
\right\rangle_{\Lambda},
\label{eq:number_excess_projected_components}
\end{equation}
where \(F_\alpha^{\mathrm{UD}}(k,t)=w_\alpha(k)F_{NN}^{\mathrm{UD}}(k,t)\). 

The transient opposition between the two projected dynamical excess contributions is quantified by
\begin{equation}
Q_C(t)
=\Delta M_N^{\mathrm{dyn},\perp}(t)
-\Delta M_N^{\mathrm{dyn},\parallel}(t),
\label{eq:tmix-direct}
\end{equation}
where \(t_C\) is the intermediate time at which their opposition is maximal. To separate the part of this opposition arising from internal redistribution from that caused by the mismatch with the structure-informed single-rate response, we define
\begin{equation}
\mathcal R(k,t)
=F_{NN}^{\mathrm{UD}}(k,t)-F_{NN}(k,t).
\end{equation}
The correction associated with each solvent-associated or solvent-orthogonal projection can then be written as
\begin{subequations}
\label{eq:number_redistribution}
\begin{align}
C_\alpha(k,t)
&=F_\alpha^{\mathrm{UD}}(k,t)-F_\alpha(k,t)\notag\\
&=w_\alpha(k)\mathcal R(k,t)+G_\alpha(k,t),
\label{eq:number_projected_correction_split}\\
G_\alpha(k,t)
&=w_\alpha(k)F_{NN}(k,t)-F_\alpha(k,t).
\end{align}
\end{subequations}
Here, \(C_\alpha\) is defined in the solvent-associated/solvent-orthogonal projection basis, whereas \(G_\alpha\) is the redistribution remaining after removal of the projected single-rate mismatch \(w_\alpha\mathcal R\). By construction, \(G_\perp=-G_\parallel\) and \(C_\perp+C_\parallel=\mathcal R\). The concentration dependence of \(G_\perp-G_\parallel\) is interpreted using the ideal-composition Bhatia--Thornton reference and its Carnahan--Starling approximation~\cite{CarnahanStarling1969,BhatiaThornton1970}. A detailed derivation is given in Sec. S7 of the SM~\cite{SupplementalMaterial}.

To characterize collective ion transport, we define the integral relaxation time of the ion-number fluctuations within each observation volume \(\Lambda\) as
\begin{equation}
T_{p,N}
=p\int_0^\infty [R_N(t)]^p\,\dd t,
\label{eq:T-p-family}
\end{equation}
where
\begin{equation}
R_N(t)=\frac{C_N(t)}{C_N(0)}
=1-\frac{M_N(t)}{2\operatorname{Var}(N_\Lambda)}.
\label{eq:R-counting}
\end{equation}
We use \(p=2\); for a single-exponential relaxation, \(T_{2,N}\) equals its decay time. Following the decomposition of \(M_N(t)\), the ideal, structure-informed single-rate, and matrix-informed responses define \(T_{2,N}^{\mathrm{id}}\), \(T_{2,N}^{\mathrm{str}}\), and \(T_{2,N}^{\mathrm{dyn}}\), respectively. For the shift analysis, \(T_{2,N}^{\mathrm{id}}\) is evaluated using the independently measured \(D_N^J\). The structural contribution is quantified by \(\Delta T_{2,N}^{\mathrm{str}}=T_{2,N}^{\mathrm{str}}-T_{2,N}^{\mathrm{id}}\), where \(T_{2,N}^{\mathrm{str}}\) additionally retains the measured \(S_{NN}(k)\). For the coupled response \(R_N^{\mathrm{dyn}}(t)=R_N^{\mathrm{str}}(t)+\delta R_N(t)\), the integral relaxation becomes
\begin{subequations}
\label{eq:T-cross-decomposition}
\begin{align}
T_{2,N}^{\mathrm{dyn}}
&=T_{2,N}^{\mathrm{str}}+T_{\mathrm{cross},N}
+T_{\mathrm{quad},N},\\
T_{\mathrm{cross},N}
&=4\int_0^\infty
R_N^{\mathrm{str}}(t)\delta R_N(t)\,\dd t,\\
T_{\mathrm{quad},N}
&=2\int_0^\infty[\delta R_N(t)]^2\,\dd t\geq0.
\end{align}
\end{subequations}
Thus, \(T_{\mathrm{cross},N}\) is signed, whereas \(T_{\mathrm{quad},N}\) is nonnegative.

The integral relaxation times are also used to estimate the collective diffusion coefficient. The structure-informed and dynamics-informed expressions are evaluated as functions of a trial diffusion coefficient \(\hat D_N\) and matched to the measured relaxation time:
\begin{equation}
T_{2,N}^{\mathrm{str}}(\hat D_N^{\mathrm{str}})
=T_{2,N}^{\Lambda},
\qquad
T_{2,N}^{\mathrm{dyn}}(\hat D_N^{\mathrm{dyn}})
=T_{2,N}^{\Lambda},
\label{eq:T2-estimator-matching}
\end{equation}
where \(T_{2,N}^{\Lambda}\) is computed from the measured \(M_N(t)\).

\section{Simulation Model and Methods}
\label{sec:methods}
This section describes the model electrolytes, molecular dynamics simulations, and calculation of the finite-volume and transport observables.

\subsection{Model electrolytes}\label{subsec:model}
We study symmetric 1:1 model electrolytes consisting of neutral LJ solvent particles and ions, all of the same size and mass \(m\)~\cite{joly2006liquid,Kim2026StructuralDynamicalCrossovers}. All solvent and ion particles interact through an LJ potential \(U_{\mathrm{LJ}}\), truncated and shifted at \(r_c=2^{1/6}\sigma\simeq1.122\,\sigma\) to retain only its repulsive part~\textcolor{black}{\cite{WeeksChandlerAndersen1971}}. For \(r<r_c\),
\begin{equation}\label{eq:lj}
U_{\mathrm{LJ}}(r)=
4\epsilon\left[
\left(\frac{\sigma}{r}\right)^{12}
-\left(\frac{\sigma}{r}\right)^6
-\left(\frac{\sigma}{r_c}\right)^{12}
+\left(\frac{\sigma}{r_c}\right)^6
\right],
\end{equation}
and \(U_{\mathrm{LJ}}(r)=0\) for \(r\geq r_c\). The same LJ energy \(\epsilon\) and diameter \(\sigma\) are used for interactions between all particle types. The Coulomb interaction \(U_C\) between ions is
\begin{equation}
U_C(r)=k_BTz_i z_j\frac{l_B}{r},
\end{equation}
where \(z_i=\pm1\) is the ionic valence and \(l_B=e^2/(4\pi\varepsilon_0\varepsilon_s k_BT)=0.2\,\sigma\) is the Bjerrum length. Here, \(\varepsilon_0\) is the vacuum permittivity and \(\varepsilon_s=5\) is the uniform background dielectric constant. The reduced ionic charges used in the simulations are \(q_i^*=z_i=\pm1\), and an asterisk denotes a quantity in reduced LJ units. The Coulomb interaction is split into short- and long-range contributions. The short-range contribution is truncated at \(3.5\,\sigma\), while the long-range contribution is calculated using the particle--particle particle--mesh (PPPM) method with a relative force accuracy of \(10^{-4}\)~\textcolor{black}{\cite{DesernoHolm1998}}.

All simulations were performed in the \(\mathrm{NpT}\) ensemble at \(p^*=p\sigma^3/\epsilon=1\) and \(T^*=k_BT/\epsilon=1\), corresponding to the liquid phase~\textcolor{black}{\cite{Kim2026StructuralDynamicalCrossovers}}. \textcolor{black}{All quantities in this work are reported in reduced LJ units unless otherwise noted.} The number of solvent particles was fixed at \(N_{\mathrm{solv}}=5000\), while the number of salt pairs \(N_{\mathrm{salt}}\) was varied with solution concentration. We define \(r_{\mathrm{salt}}=N_{\mathrm{salt}}/N_{\mathrm{solv}}\) and \(x_{\mathrm{salt}}=N_{\mathrm{ion}}/(N_{\mathrm{ion}}+N_{\mathrm{solv}})=2r_{\mathrm{salt}}/(1+2r_{\mathrm{salt}})\), with \(N_{\mathrm{ion}}=2N_{\mathrm{salt}}\). The six concentrations studied are \(r_{\mathrm{salt}}=0.01,0.02,0.05,0.10,0.20,\) and \(0.40\). The equations of motion were integrated using the velocity--Verlet algorithm with a timestep \(\delta t^*=\delta t\sqrt{\epsilon/(m\sigma^2)}=0.001\). The target temperature and pressure were maintained using the Nos\'e--Hoover thermostat and barostat with damping times of \(0.1\) and \(1.0\) LJ time units, respectively, corresponding to 100 and 1000 integration steps~\textcolor{black}{\cite{Nose1984,Hoover1985,MartynaTobiasKlein1994}}. For each condition, we obtained ten independent long trajectories for static structure and transport and an additional ten high-frequency trajectories for the short-time finite-volume fluctuations. The solvent-resolved matrices use five or ten matched trajectories depending on the state.

\begin{figure*}[t]
\centering
\includegraphics[width=0.80\textwidth]{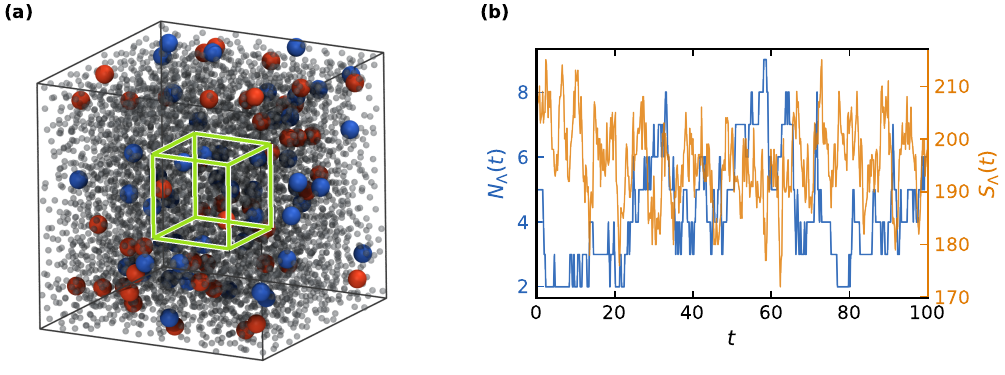}
\caption{Illustration of finite-volume counting and the measured count trajectories. (a) Simulation snapshot at \(r_{\mathrm{salt}}=0.01\), showing a cubic observation box with \(L_{\mathrm{obs}}=8\). The displayed particle sizes are not to scale. (b) Time series of the ion-number count \(N_\Lambda(t)\) and solvent-number count \(S_\Lambda(t)\) within the same observation box.}
\label{fig:snapshot}
\end{figure*}

\subsection{Finite-volume observables and diffusion estimation}

We evaluated finite-volume counts in nonoverlapping cubic observation volumes with side lengths \(L_{\mathrm{obs}}\) ranging from 2 to 12. The cubes were fixed in physical coordinates and separated by approximately \(\sigma\) (see Fig.~\ref{fig:snapshot} for an observation volume and its ion and solvent count trajectories). The continuum window average \(\avg{A}_{\Lambda}\) in Eq.~\eqref{eq:window_average} was implemented as the \(1/V\)-normalized sum over nonzero reciprocal vectors, retaining the exact cubic kernel \(f_\Lambda(\vect k)\) on every allowed mode. Reciprocal-space correlation functions were evaluated using instantaneous fractional coordinates on the mean-cell grid to remove affine cell dilation, extending the reference-cell convention used for static structure~\cite{Cheng2022}. The leading volume-fluctuation contribution in the \(\mathrm{NpT}\) ensemble was removed from the count dynamics by an a posteriori correction. The posterior correction was compared with configuration-wise subtraction, and both count and reciprocal-space observables were checked against fixed-volume \(\mathrm{NVT}\) controls (see Sec.~S1 of the SM~\cite{SupplementalMaterial}). For the integral relaxation-time analysis [Eq.~\eqref{eq:T-p-family}], posterior volume corrections were applied to both \(M_N(t)\) and \(\operatorname{Var}(N_\Lambda)\) in \(R_N(t)=1-M_N(t)/[2\operatorname{Var}(N_\Lambda)]\), as described in Sec.~S1 of the SM~\cite{SupplementalMaterial}. A common finite-\(N\) correction factor cancels in this normalized relaxation.

The independent collective diffusion coefficient \(D_N^J\) was obtained~\cite{Kim2026StructuralDynamicalCrossovers} using the Einstein--Helfand relation~\cite{Helfand1960},
\begin{equation}
D_N^J=\frac{1}{6N_{\mathrm{ion}}}\lim_{t\rightarrow\infty}\frac{\mathrm{d}}{\mathrm{d}t}\avg{|\Delta\vect R_N(t)|^2},
\label{eq:DXJ}
\end{equation}
where \(\Delta\vect R_N(t)=\sum_{i\in\mathrm{ions}}\Delta\vect r_i(t)\) is the collective ion displacement. We fitted the slope over \(150\leq t\leq250\); representative long-time plateaus and fitting-window convergence tests are reported in Sec.~S4 of the SM~\cite{SupplementalMaterial}. The tagged-particle diffusion coefficient \(D_{\mathrm{self}}\) was obtained from the ion-averaged MSD, which contains only single-particle displacement correlations~\cite{Kim2026StructuralDynamicalCrossovers}.

The direct finite-window estimate \(\hat D_N^{\mathrm{id}}(L_{\mathrm{obs}})\) was obtained by fitting Eq.~\eqref{eq:number_ideal_analytic} to the posterior-volume-corrected \(M_N(t)\) over \(0<t\leq t_C\), with \(\tau_v=D\) in reduced LJ units. This procedure was used for all direct-fit estimates in Fig.~\ref{fig:integral-D-six}, without applying a finite-\(N\) correction to the measured MSF (see Sec.~S1 of the SM~\cite{SupplementalMaterial}). The endpoint \(t_C\) was evaluated separately for each trajectory and observation length. The one-parameter objective was
\begin{equation}
\mathcal L(D)=\frac{1}{n_b}\sum_{i=1}^{n_b}
\left[
\ln M_{N}(t_i)
-\ln M_N^{\mathrm{id}}(t_i;D)
\right]^2,
\label{eq:direct-fit-objective}
\end{equation}
Before fitting, the data were averaged over a total of \(n_b\) logarithmically spaced time bins, and the resulting estimates were averaged over independent trajectories (see Sec.~S4 in the SM~\cite{SupplementalMaterial} for the direct-fit analysis).

\section{Results and Discussion}
This section examines finite-volume ion-number fluctuations in the symmetric
1:1 explicit-solvent model across six salt concentrations. We first establish
that the ion-number response at early times appears nearly ideal and then resolve the structural and dynamical
contributions hidden within it. We next determine how these contributions
affect finite-volume transport estimates and use wave-vector-resolved
number--solvent correlations to identify the role of ion--solvent coupling.

\begin{figure*}[t]
\centering
\includegraphics[width=0.98\textwidth]{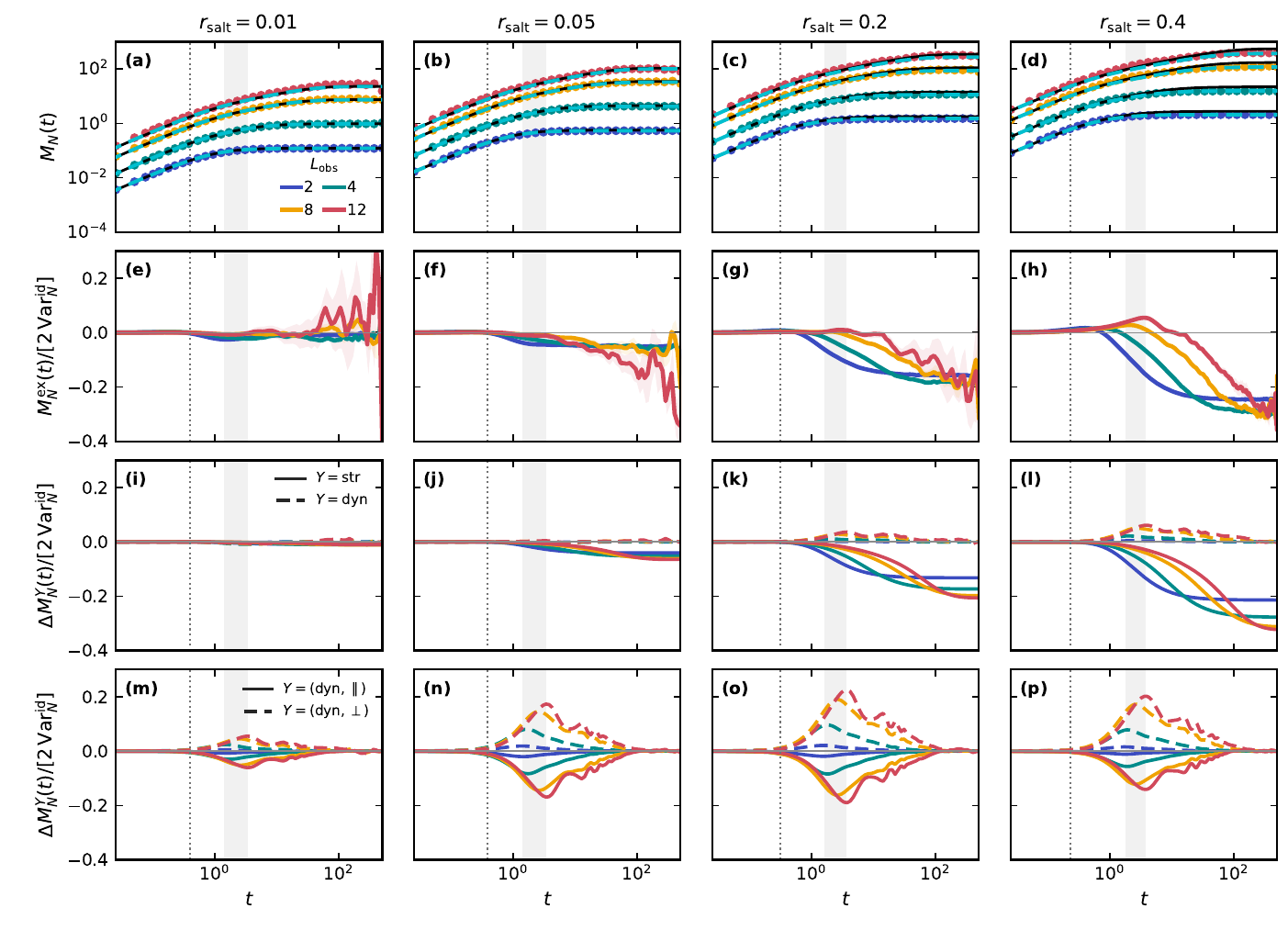}
\caption{Ion-number MSF \(M_N(t)\) and its resolved nonideal contributions.
(a--d) Computed \(M_N(t)\) (markers) [Eq.~\eqref{eq:exact_msf_bulk}] at four salt concentrations \(r_{\mathrm{salt}}\). Black solid and cyan dashed curves denote the ideal [Eq.~\eqref{eq:number_ideal_analytic}] and structure-informed single-rate predictions [Eqs.~\eqref{eq:exact_msf_bulk} and \eqref{eq:inertial_phi}], respectively, evaluated using the independent \(D_N^J\) [Eq.~\eqref{eq:DXJ}].
(e--h) Total excess \(M_N^{\mathrm{ex}}(t)\) [Eq.~\eqref{eq:number_excess_all_components}].
(i--l) Structural (solid) and dynamical (dashed) excess contributions [Eq.~\eqref{eq:number_excess_component_definitions}].
(m--p) Solvent-associated (solid) and solvent-orthogonal (dashed) components of the dynamical excess [Eq.~\eqref{eq:number_excess_projected_components}]. Each column represents one salt concentration, and colors denote \(L_{\mathrm{obs}}=2,4,8,12\). All excess contributions are normalized by the ideal long-time plateau, \(2\,\operatorname{Var}^{\mathrm{id}}(N_\Lambda)\). Vertical dotted lines indicate \(\tau_v\), and gray bands indicate the range of \(t_C\) [Eq.~\eqref{eq:tmix-direct}]. Additional details are given in Secs.~S2 and S6 of the SM~\cite{SupplementalMaterial}.}
\label{fig:numberMSF}
\end{figure*}

\subsection{Ion-number fluctuations appear nearly ideal despite resolved nonideal contributions}
\label{sec:results-number}

Figure~\ref{fig:numberMSF} shows the evolution of the ion-number MSF, \(M_N(t)\) and its resolved contributions, over several salt concentrations and observation lengths. The relaxation passes through three stages. At \(t\ll\tau_v\), ballistic crossings of the sharp observation boundary give \(M_N(t)\propto t\) [Eq.~\eqref{eq:sharp-window-ballistic}]. The leading coefficient is independent of \(D_N^J\), so this regime contains little information about collective diffusion. For \(t\gtrsim\tau_v\), boundary crossings become sensitive to collective transport, while the local exponent \(\alpha_N(t)=d\ln M_N(t)/d\ln t\) decreases continuously rather than forming an extended \(M_N(t)\propto t^{1/2}\) regime. \textcolor{black}{The corresponding local-exponent curves are shown in Fig.~S3.1 of the SM~\cite{SupplementalMaterial}.} This behavior contrasts with the overdamped colloidal regime studied previously, in which the square-root amplitude provides a direct diffusion estimate~\cite{Mackay2024,Carter2025}. The difference is consistent with the limited separation among velocity relaxation, ion--solvent collective relaxation, and finite-volume saturation in the present model. Finally, \(M_N(t)\) approaches the long-time plateau \(2\,\operatorname{Var}(N_\Lambda)\), which is determined by $S_{NN}(k)$.

The computed \(M_N(t)\) curves in Figs.~\ref{fig:numberMSF}(a)--(d) appear close to the independent-particle inertial reference, \(M_N^{\mathrm{id}}(t)\) [Eq.~\eqref{eq:number_ideal_analytic}], even as salt concentration and observation length increase. However, as shown in Fig.~\ref{fig:integral-D-six}, fitting this reference to the computed \(M_N(t)\) does not consistently recover the independently measured collective diffusion coefficient \(D_N^J\), implying the importance of seemingly small nonideal contributions. The decomposition of \(M_N(t)\) clearly shows that this apparent ideality does not imply the absence of correlations (Fig.~\ref{fig:numberMSF}). We separate the measured excess \(M_N^{\mathrm{ex}}(t)\) (Figs.~\ref{fig:numberMSF}(e)--(h)) into a structural contribution \(\Delta M_N^{\mathrm{str}}(t)\)  (Eq.~\eqref{eq:number_excess_component_definitions}) obtained by introducing the measured \(S_{NN}(k)\) into both the static amplitude and the structure-informed single-rate relaxation, and a dynamical contribution  \(\Delta M_N^{\mathrm{dyn}}(t)\)  (Eq.~\eqref{eq:number_excess_component_definitions}) measuring the departure of the measured \(F_{NN}(k,t)\) from this single-rate approximation. 

As expected from the sharp-window short-time limit, no excess contribution is resolved in the ballistic regime below \(\tau_v\) at all the examined salt concentrations. This does not imply that structural and dynamical correlations are absent; rather, they do not enter the leading surface-crossing term and appear only at higher orders (Eq.~\ref{eq:sharp-window-ballistic}). The total excess \(M_N^{\mathrm{ex}}(t)\) remains small over the transport-sensitive crossover interval because its structural and dynamical contributions partially compensate (Figs.~\ref{fig:numberMSF}(i)--(l)), and its magnitude grows as \(M_N(t)\) approaches the long-time plateau. In particular, the increasingly negative \(\Delta M_N^{\mathrm{str}}(t)\) reflects the suppression of long-wavelength ion-number fluctuations, \(S_{NN}(k\rightarrow0)<1\), consistent with the low compressibility and packing and composition constraints of the dense mixture. Its magnitude generally increases with \(L_{\mathrm{obs}}\) and \(r_{\mathrm{salt}}\), as shown directly in Figs.~\ref{fig:numberMSF}(i)--(l). Nonetheless, the long-time plateau lies outside the direct fitting interval at the early times and therefore does not enter that inference directly, although the same structural correlations affect the transient response and the integral relaxation estimator. The full decomposition, including \(\Delta M_N^{\mathrm{res}}(t)\), is shown in Fig.~S6.3 of the SM~\cite{SupplementalMaterial}.

Over the intermediate crossover interval between the ballistic and plateau regimes, the positive dynamical excess \(\Delta M_N^{\mathrm{dyn}}(t)\) opposes the negative structural excess \(\Delta M_N^{\mathrm{str}}(t)\). The positive \(\Delta M_N^{\mathrm{dyn}}(t)\) indicates that the measured finite-window ion-number correlation is lower than the structure-informed single-rate prediction based on \(D_N^J\) [Eq.~\eqref{eq:number_excess_component_definitions}]. At $r_{\mathrm{salt}}=0.4$, \(M_N^{\mathrm{ex}}(t)\) even becomes positive for large observation volumes in the same time range as the weak shoulder in \(M_N(t)\) (Figs.~\ref{fig:numberMSF}(d) and \ref{fig:numberMSF}(h)). We find that the relatively small \(\Delta M_N^{\mathrm{dyn}}(t)\) is the remainder of two larger, compensating contributions: the solvent-associated (\(\Delta M_N^{\mathrm{dyn},\parallel}(t)\)) and solvent-orthogonal (\(\Delta M_N^{\mathrm{dyn},\perp}(t)\)) terms [Eq.~\eqref{eq:number_excess_projected_components}]. Their opposition reveals an internal redistribution in the transport-sensitive crossover regime that is largely hidden in the total \(M_N(t)\) (Figs.~\ref{fig:numberMSF}(m)--(p)). Their opposition develops after \(\tau_v\), reaches its largest contrast at \(t_C\), and weakens as \(M_N(t)\) approaches its plateau. We therefore define \(t_C\) as the time at which \(\Delta M_N^{\mathrm{dyn},\perp}(t)-\Delta M_N^{\mathrm{dyn},\parallel}(t)\) is maximal [Eq.~\eqref{eq:tmix-direct}]. This operational crossover time identifies the most clearly resolved internal redistribution and defines the upper bound of the direct fitting interval used below; however, it is not interpreted as an independent transport timescale.

\begin{figure*}[t]
\centering
\includegraphics[width=0.98\textwidth]{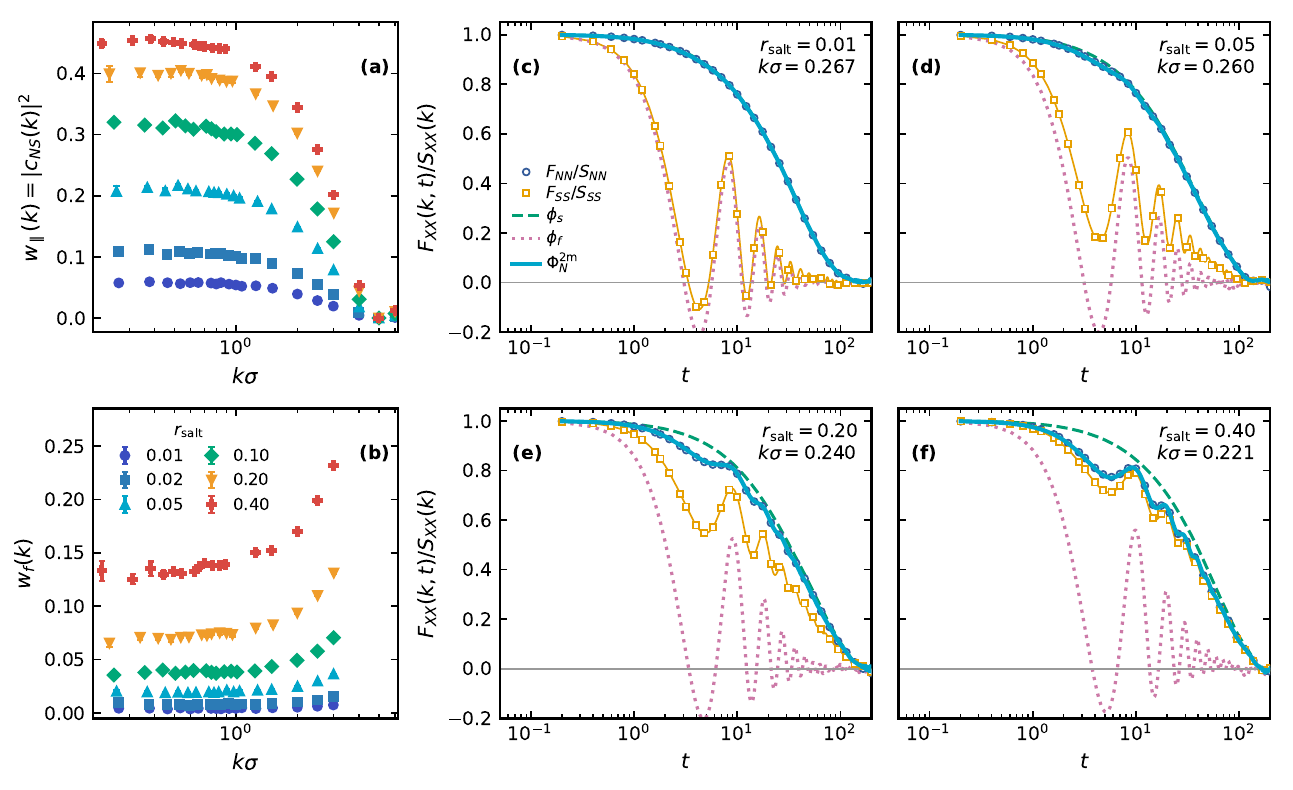}
\caption{Static and dynamical measures of ion--solvent coupling.
(a) Equal-time solvent-associated fraction \(w_{\parallel}(k)=|c_{NS}(k)|^2\) for all six concentrations [Eq.~\eqref{eq:even_static_matrix}].
(b) Fitted weight \(w_f(k)\) of the fast-like component in the statically whitened \(N\)--\(S\) relaxation matrix [Eq.~\eqref{eq:NS_couple}].
(c--f) Normalized intermediate scattering functions at the lowest sampled wave number for \(r_{\mathrm{salt}}=0.01,0.05,0.20,0.40\). Blue circles and orange squares denote the computed \(F_{NN}(k,t)/S_{NN}(k)\) and \(F_{SS}(k,t)/S_{SS}(k)\), respectively. Green dashed and magenta dotted curves denote the slow-like and fast-like spectral components, and the cyan solid curve is their two-component reconstruction, \(\Phi_N^{2\mathrm m}=(1-w_f)\phi_s+w_f\phi_f\). \textcolor{black}{The measured matrix elements and two-mode construction are detailed in Secs.~S6 and S8 of the SM, respectively~\cite{SupplementalMaterial}.}}
\label{fig:all-six-matrix-interference}
\end{figure*}

\subsection{Ion--solvent coupling reveals hidden dynamical redistribution and its crossover}
\label{sec:results-fkt}
To identify the wave-vector-dependent origin of this finite-volume redistribution, we examine the static and dynamical mixing between ion-number and solvent fluctuations. The equal-time projection \(w_{\parallel}(k)=|c_{NS}(k)|^2\) (Eq.~\eqref{eq:even_static_matrix}), which measures the static visibility of the solvent-associated component, and the fitted weight \(w_f(k)\) (Eq.~\eqref{eq:NS_couple}), which measures the contribution of the fast-like spectral component in the ion-number relaxation, are shown in Figs.~\ref{fig:all-six-matrix-interference}(a)--(b). Both increase with \(r_{\mathrm{salt}}\) and exhibit similar wave-number-dependent trends, consistent with the increasing \(\Delta M_N^{\mathrm{dyn}}(t)\).

The measured reciprocal-space correlators clearly resolve ion--solvent mixing at high salt concentrations through a slow ion-number-dominated relaxation and a faster, oscillatory solvent-rich relaxation (Figs.~\ref{fig:all-six-matrix-interference}(c)--(f)). 
\begin{equation}\label{eq:NS_couple}
\frac{F_{NN}(k,t)}{S_{NN}(k)}
\simeq
[1-w_f(k)]\phi_s(k,t)
+w_f(k)\phi_f(k,t).
\end{equation}
Here, \(\phi_s(k,t)\) denotes the slow ion-number-rich component, whereas \(\phi_f(k,t)\) denotes the faster solvent-rich component.
At \(r_{\mathrm{salt}}=0.01\), the measured ion-number and solvent responses are nearly decoupled over the sampled wave-vector and time ranges. As \(r_{\mathrm{salt}}\) increases, the faster solvent-rich component \(\phi_f(k,t)\) acquires a finite projection onto the ion-number response. This component is most clearly resolved at low wave vectors, whereas its projection onto the ion-number response becomes weak and eventually unresolved by \(k\sigma\simeq1\)--\(1.5\). Its weak projection onto the ion-number response is consistent with the shoulder observed at high concentration (Fig.~\ref{fig:numberMSF}(d)), although this correspondence does not identify a unique microscopic origin for the shoulder. The fitted components describe the measured relaxation of $F_{NN}(k,t)/S_{NN}(k)$ but are not assumed to be time-independent kinetic normal modes. Their distinct weights and relaxation rates produce an observation-length-dependent response in \(M_N(t)\) because a finite observation volume samples a range of wave vectors. 

\begin{figure}[!t]
\centering
\includegraphics[width=0.78\columnwidth]{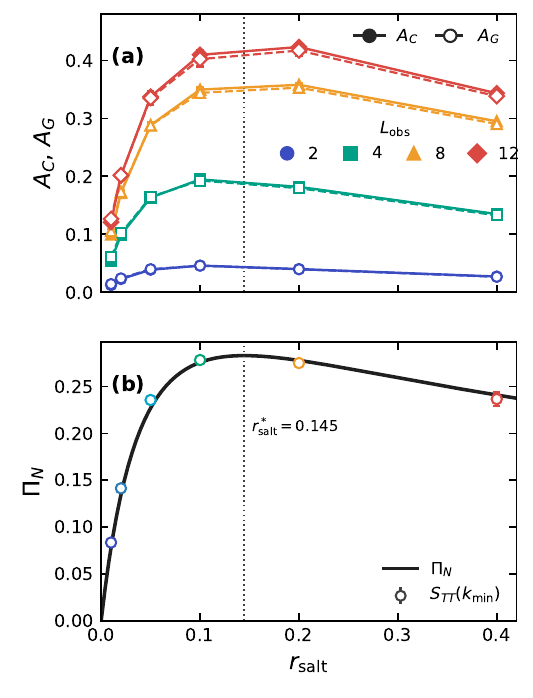}
\caption{Intermediate-concentration turnover of the projected dynamical redistribution.
(a) Computed opposition \(A_C\) [Eq.~\eqref{eq:redistribution-contrast-amplitudes}] (filled symbols) and its closure-free counterpart \(A_G\) (open symbols) for \(L_{\mathrm{obs}}=2,4,8,\) and \(12\).
(b) Static BT projection factor \(\Pi_N\) [Eq.~\eqref{eq:bt-redistribution-contrast}] obtained using the ideal-composition/CS approximation (line) and the measured \(S_{TT}(k_{\min})\) (symbols). Vertical dotted lines mark the maximum predicted by the CS approximation. Details of this analysis are given in Sec.~S7 of the SM~\cite{SupplementalMaterial}.}
\label{fig:redistribution-turnover}
\end{figure}

It is notable that the maximal opposition \(A_C\) of the two projected contributions is nonmonotonic with \(r_{\mathrm{salt}}\), despite the monotonic increase of both \(w_\parallel(k)\) and \(w_f(k)\) (Fig.~\ref{fig:redistribution-turnover}).
\begin{equation}
A_C=\frac{Q_C(t_C)}{2\operatorname{Var}^{\mathrm{id}}(N_\Lambda)}.
\label{eq:redistribution-contrast-amplitudes}
\end{equation}
Again, $t_C=\operatorname*{arg\,max}_{t}Q_C(t)$. The turnover behavior of \(A_C\) motivates a comparison with the Bhatia--Thornton (BT) reference~\cite{BhatiaThornton1970}, which expresses the ion and solvent fields in terms of total-density ($T$) and composition ($c$) fluctuations (see Sec.~S7 in the SM~\cite{SupplementalMaterial} for the detailed derivation). 
Because \(A_C\) also contains the mismatch with the structure-informed single-rate response, we first define \(A_G\) as its closure-free counterpart. 
\[
A_G=\frac{|Q_G(t_G)|}{2\operatorname{Var}^{\mathrm{id}}(N_\Lambda)},
\]
where $t_G=\operatorname*{arg\,max}_{t}|Q_G(t)|$, and \(Q_G\) is the finite-volume projection of \(G_\perp-G_\parallel\), describing the closure-free redistribution: \(Q_C(t)=Q_G(t)+Q_R(t)\). This decomposition follows from Eq.~\eqref{eq:number_projected_correction_split}. We note that \(Q_R\) is the projected contribution of the single-rate mismatch. Both \(A_C\) and \(A_G\) increase from the dilute regime, are largest over \(r_{\mathrm{salt}}=0.10\)--\(0.20\), and decrease at \(r_{\mathrm{salt}}=0.40\) (Fig.~\ref{fig:redistribution-turnover}). The persistence of the turnover in \(A_G\) shows that it is not produced solely by the single-rate mismatch. Moreover, \(t_G\) agrees with \(t_C\) within the native time resolution, indicating that the two amplitudes identify the same redistribution transient.

The \(k\)-resolved closure-free contrast is factorized under the ideal-composition  (IC) assumptions \(S_{Tc}=0\) and \(S_{cc}=x_{\mathrm{salt}}(1-x_{\mathrm{salt}})\)\textcolor{black}{, with independently relaxing total-density and composition fluctuations,} as
\begin{equation}
\begin{aligned}
G_\perp^{\mathrm{IC}}(k,t)-G_\parallel^{\mathrm{IC}}(k,t)
&=2\Pi_N(x_{\mathrm{salt}};S_{TT})\\
&\quad\times[\phi_c(k,t)-\phi_T(k,t)].
\end{aligned}
\label{eq:bt-redistribution-contrast}
\end{equation}
Here, \(\Pi_N\) is the static projection factor that determines how strongly the contrast between the total-density and composition relaxations appears in the ion-number response, whereas \(\phi_c-\phi_T\) is their dynamical contrast. Using the Carnahan--Starling approximation for the packing-suppressed \(S_{TT}(0)\), \(\Pi_N\) is maximal between \(r_{\mathrm{salt}}=0.10\) and \(0.20\). After one vertical rescaling at each observation length, this prediction follows the concentration dependence of both \(A_C\) and \(A_G\) (Fig.~\ref{fig:redistribution-turnover}). The turnover therefore does not require ion--solvent mixing to weaken at high concentration. Instead, the increasing ion fraction initially strengthens the projection, while the decreasing solvent fraction eventually reduces the simultaneous projection of the two BT fields. The BT/CS reference thus rationalizes the intermediate-concentration turnover of the redistribution, whereas the measured relaxation matrix determines its absolute magnitude and timing.

The ideal-composition BT/CS approximation above rationalizes the concentration turnover of the redistribution amplitude \(A_C\), but it does not determine the time \(t_C\) at which the opposition is maximal. We analyze \(t_C\) separately using a reciprocal-space counterpart based on the coupled two-mode approximation [Eq.~\eqref{eq:NS_couple}]. Using the slow-like and fast-like instantaneous spectral components, \(\phi_s(k,t)\) and \(\phi_f(k,t)\), of the statically whitened \(N\)--\(S\) relaxation matrix, we define
\begin{equation}
t_C^{2m}(L_{\mathrm{obs}})=\underset{t}{\operatorname{arg\,max}}\,Q_C^{2m}(t),
\label{eq:tmix-two-mode-argmax}
\end{equation}
where\[Q_C^{2m}(t)=\avg{S_{NN}(k)w_f(k)[\phi_s(k,t)-\phi_f(k,t)]}_{\Lambda},\] and \(w_f(k)\) is the fitted nonnegative weight of the fast-like component. The construction reduces to a single-response limit as \(w_f\to0\). The resulting \(t_C^{2m}\) closely follows \(t_C\) over most conditions examined. \textcolor{black}{This agreement supports interpreting \(t_C\) through the dynamical contrast between the slow ion-number-rich and fast solvent-rich relaxation components. Thus, \(t_C^{2m}\) provides a reciprocal-space estimate} of the fitting endpoint for \(M_N(t)\), without requiring the projected finite-window excess contributions to be constructed separately. Details are given in Sec.~S8 of the SM~\cite{SupplementalMaterial}.

\subsection{finite-volume relaxation yields scale-dependent estimates of collective ion diffusion}
\label{sec:results-transport}

\begin{figure*}[t]
\centering
\includegraphics[width=0.80\textwidth]{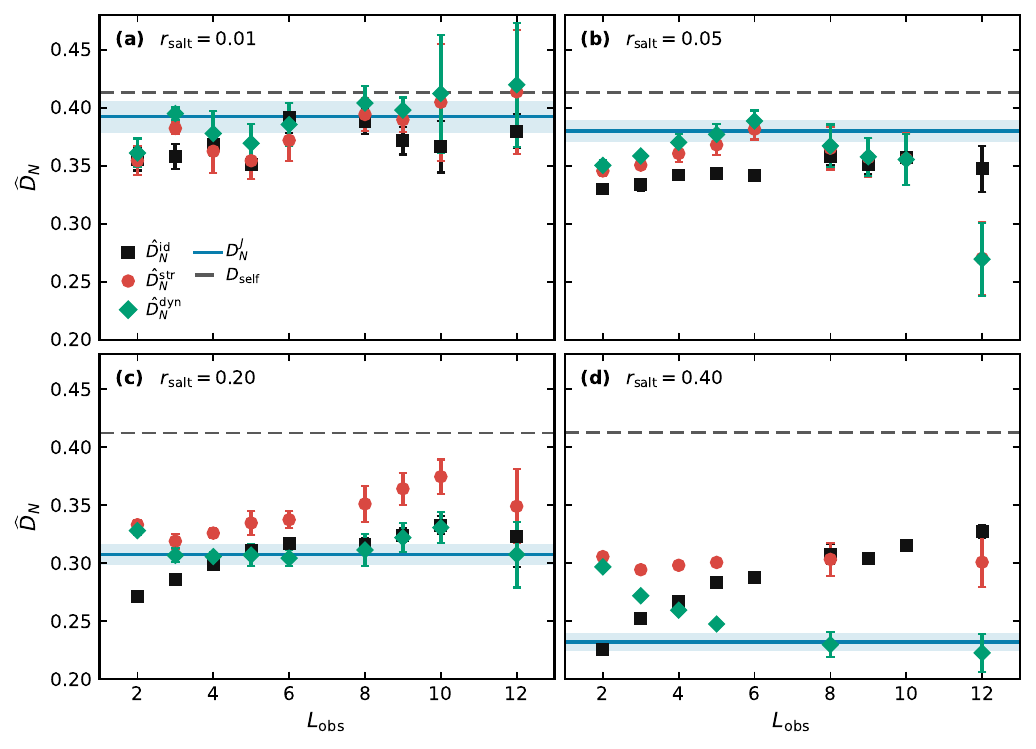}
\caption{Scale-dependent estimates of collective ion diffusion from finite-volume relaxation.
(a--d) Estimates of \(\hat D_N\) at \(r_{\mathrm{salt}}=0.01,0.05,0.20,\) and \(0.40\). Black squares denote the direct ideal-MSF estimate \(\hat D_N^{\mathrm{id}}\), obtained by fitting Eq.~\eqref{eq:number_ideal_analytic} over \(0<t\leq t_C\). Red circles and green diamonds denote the structure-informed single-rate estimate \(\hat D_N^{\mathrm{str}}\) and coupled \(N\)--\(S\) estimate \(\hat D_N^{\mathrm{dyn}}\), respectively, obtained by matching the corresponding \(T_{2,N}(\hat D)\) to the computed \(T_{2,N}^{\Lambda}\) [Eq.~\eqref{eq:T2-estimator-matching}]. Blue bands and gray dashed lines denote the independently computed \(D_N^J\) [Eq.~\eqref{eq:DXJ}] and \(D_{\mathrm{self}}\), respectively. Objective-profile analysis of \(\hat D_N^{\mathrm{id}}\) is given in Sec.~S4, and details of the coupled-relaxation analysis are given in Sec.~S8 of the SM~\cite{SupplementalMaterial}.}
\label{fig:integral-D-six}
\end{figure*}

We now examine how these structural and dynamical correlations affect the collective diffusion coefficient estimated from finite-volume relaxation. Figure~\ref{fig:integral-D-six} compares three estimators of \(D_N\). The first, \(\hat{D}_N^{\mathrm{id}}\), is obtained by directly fitting \(M_N^{\mathrm{id}}(t;\hat{D})\) to the computed \(M_N(t)\). The other two incorporate correlations through the integral relaxation time \(T_{2,N}(\hat{D})\) [Eq.~\eqref{eq:T-p-family}]: the structure-informed single-rate estimator \(\hat{D}_N^{\mathrm{str}}\) uses the measured \(S_{NN}(k)\), whereas the coupled estimator \(\hat{D}_N^{\mathrm{dyn}}\) additionally uses the matrix-informed \(\Phi_N(k,t)\) [Eq.~\eqref{eq:NS_couple}]. In the latter two cases, \(\hat{D}\) is determined by matching the model \(T_{2,N}(\hat{D})\) to \(T_{2,N}^{\Lambda}\) computed from \(M_N(t)\) [Eq.~\eqref{eq:T2-estimator-matching}]. The independently measured collective diffusion coefficient \(D_N^J\) [Eq.~\eqref{eq:DXJ}] and tagged-particle self-diffusion coefficient \(D_{\mathrm{self}}\) are also shown as references. The measured collective-to-self diffusion ratio is compatible with \(D_N^J/D_{\mathrm{self}}=1-x_{\mathrm{salt}}\). This relation follows from ideal label statistics under the zero-total-momentum constraint and independently supports the ideal-composition reference used to interpret \(\Phi_N(k,t)\), although it does not determine the measured two-channel dynamics (see Sec.~S7 in the SM~\cite{SupplementalMaterial}). We note that the near-ideal early-time curve does not guarantee an accurate \(\hat{D}_N^{\mathrm{id}}\), because its leading ballistic term is independent of \(D\). The estimate is instead determined by the weaker pre-crossover curvature and is therefore sensitive to small excess contributions (see Sec.~S4 in the SM~\cite{SupplementalMaterial} for the objective-profile analysis).

Across most cases in Fig.~\ref{fig:integral-D-six}, the three estimators lie closer to \(D_N^J\) than to \(D_{\mathrm{self}}\), consistent with the collective nature of the finite-volume relaxation. The accuracy of the ideal estimator \(\hat{D}_N^{\mathrm{id}}(L_{\mathrm{obs}})\) depends on observation length and varies nonmonotonically with salt concentration. As discussed above, its accuracy does not indicate the absence of correlations but can instead result from compensation between the structural and dynamical excess contributions. At the lowest concentration examined, \(r_{\mathrm{salt}}=0.01\), \(\hat{D}_N^{\mathrm{id}}\) gives a comparatively accurate estimate of \(D_N^J\) over most observation lengths. Its accuracy deteriorates at \(r_{\mathrm{salt}}=0.05\), but improves again at \(r_{\mathrm{salt}}=0.20\). The agreement at \(r_{\mathrm{salt}}=0.20\) is consistent with partial compensation in the total \(M_N(t)\) between the negative structural excess and positive dynamical excess, leaving only a small deviation. At this concentration, including structural correlations alone does not improve the overall agreement with \(D_N^J\), whereas accounting for the coupled ion--solvent relaxation improves it. The significance of the coupled relaxation is most clearly demonstrated at \(r_{\mathrm{salt}}=0.40\), where both \(\hat{D}_N^{\mathrm{id}}\) and \(\hat D_N^{\mathrm{str}}\) deviate from \(D_N^J\), whereas \(\hat{D}_N^{\mathrm{dyn}}\) substantially reduces the discrepancy.

\begin{figure}[!t]
\centering
\includegraphics[width=0.78\columnwidth]{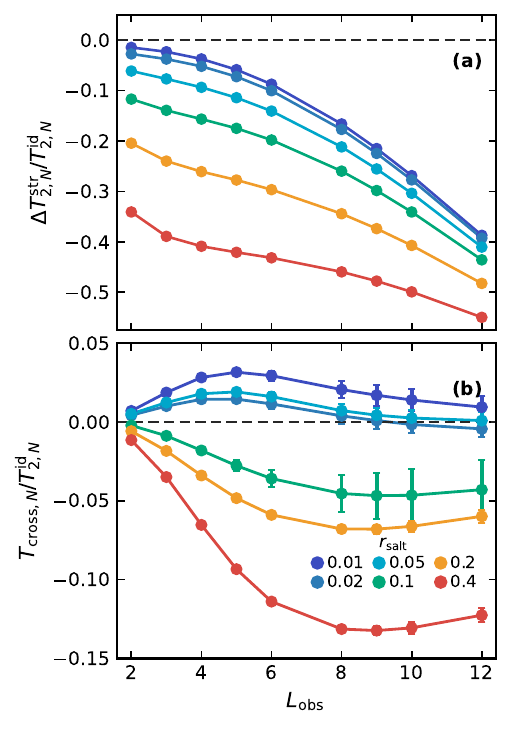}
\caption{Effects of static structure and coupled dynamics on the integral relaxation.
(a) Relative structural change \(\Delta T_{2,N}^{\mathrm{str}}/T_{2,N}^{\mathrm{id}}\).
(b) Normalized cross contribution \(T_{\mathrm{cross},N}/T_{2,N}^{\mathrm{id}}\) arising from the coupled ion--solvent relaxation [Eq.~\eqref{eq:T-cross-decomposition}]. Colors denote the six salt concentrations, and horizontal dashed lines indicate zero. Both quantities are normalized by the same ideal relaxation time \(T_{2,N}^{\mathrm{id}}\) [Eq.~\eqref{eq:T-p-family}].}
\label{fig:DL-integral}
\end{figure}

Figure~\ref{fig:DL-integral} shows how structural and dynamical contributions shift the integral relaxation times, using the common normalization \(T_{2,N}^{\mathrm{id}}\). The structure-informed single-rate response gives \(\Delta T_{2,N}^{\mathrm{str}}<0\) at all concentrations, and its magnitude generally increases with \(r_{\mathrm{salt}}\) and \(L_{\mathrm{obs}}\) (Fig.~\ref{fig:DL-integral}(a)). The cross contribution from coupled ion--solvent relaxation exhibits a different concentration dependence (Fig.~\ref{fig:DL-integral}(b)). It is small and positive in the dilute regime but becomes increasingly negative for \(r_{\mathrm{salt}}\geq0.10\). Thus, the cross contribution weakly opposes the structural shift at low concentration but acts in the same direction at intermediate and high concentrations.

This sign change follows from the dynamical MSF excess. For \(R_N^{\mathrm{dyn}}(t)=R_N^{\mathrm{str}}(t)+\delta R_N(t)\),
\begin{equation}
\delta R_N(t)=-\frac{\Delta M_N^{\mathrm{dyn}}(t)}
{2\operatorname{Var}(N_\Lambda)},
\label{eq:matrix-relaxation-shift}
\end{equation}
which enters the cross term \(T_{\mathrm{cross},N}\) in Eq.~\eqref{eq:T-cross-decomposition}. At intermediate and high concentrations, \(\Delta M_N^{\mathrm{dyn}}(t)>0\) over the dominant relaxation interval, giving \(\delta R_N(t)<0\) and hence \(T_{\mathrm{cross},N}<0\). In the dilute regime, the smaller dynamical excess changes sign and gives a weakly positive cross term. The associated quadratic term, \(2\int_0^\infty[\delta R_N(t)]^2\,\mathrm{d}t\), is nonnegative but smaller over the conditions examined.

This result also clarifies why the turnover of \(A_C\) does not directly determine the transport correction. \(A_C\) measures the opposition between the solvent-associated and solvent-orthogonal contributions, whereas \(T_{\mathrm{cross},N}\) depends on their remaining sum and its overlap with \(R_N^{\mathrm{str}}(t)\). A large \(A_C\) can therefore coexist with a small coupled-dynamics correction when the two projected contributions cancel strongly. Conversely, less complete cancellation at high concentration can produce a larger coupled-dynamics correction even as \(A_C\) decreases. This concentration-dependent balance between structure and coupled ion--solvent dynamics explains why the accuracy of the finite-volume estimators depends on both \(r_{\mathrm{salt}}\) and \(L_{\mathrm{obs}}\).

\section{Conclusions}
In this work, we formulated and tested an explicit-solvent extension of finite-volume fluctuation analysis using a symmetric 1:1 solvent primitive model. In this model, cation--anion exchange symmetry allows ion-number--solvent coupling while forcing charge cross correlations to vanish within the two-point structure and relaxation matrices. The finite-volume ion-number response appears close to the independent-particle form, but we show that this apparent ideality reflects compensation among structural and dynamical contributions. The number--solvent relaxation further reveals an internal redistribution between solvent-associated and solvent-orthogonal projections whose amplitude is largest at intermediate concentration. An ideal-composition Bhatia--Thornton reference rationalizes this turnover through the concentration dependence of the static projection, while the measured dynamics determine its magnitude and timing.

These structural and dynamical correlations affect the collective ion-diffusion coefficient estimated from finite-volume relaxation despite the apparent ideality of \(M_N(t)\). The resulting estimate depends on \(L_{\mathrm{obs}}\) and on whether the response is described by an ideal, structure-informed single-rate, or coupled-field closure. Structure alone does not systematically improve the estimate, whereas including the coupled number--solvent relaxation substantially reduces the high-concentration discrepancy from \(D_N^J\). Thus, close agreement of \(M_N(t)\) with the ideal reference does not guarantee an accurate diffusion estimate, while its dependence on \(L_{\mathrm{obs}}\) reflects finite-wave-vector relaxation and sensitivity to the assumed closure.

Finite-volume fluctuations provide a microscopic-to-macroscopic bridge by retaining particle-level correlations while testing the density-field closure used to infer collective transport~\cite{Dean1996,Illien2025}. The present symmetric model provides a controlled reference in which ion-number--solvent coupling is separated from charge fluctuations, although it does not independently distinguish electrostatic effects from packing and ideal-composition statistics. Extending the analysis to the full \(N\)--\(Z\)--\(S\) structure and relaxation matrix could establish how asymmetry, screening, and specific solvation jointly control collective transport in more realistic electrolytes~\cite{CiachGozdzStell2007,McDanielSon2018,IllienCarofRotenberg2024,Minh2026Asymmetric}. With solvent-polarization variables included for polar solvents, this framework could also connect finite-volume fluctuations to SDFT descriptions of coupled charge--polarization relaxation, solvent-induced memory, and nonlocal dielectric effects on ionic transport~\cite{Dean1996,Illien2025,IllienCarofRotenberg2024,VargheseRotenbergIllien2026,DemeryToquer2026Solvent}.

\section*{Supplemental Material}
The Supplemental Material contains \(\mathrm{NpT}\)-ensemble corrections and controls, low-wave-number structural analysis, tests of the diffusion estimates, and details of the ion--solvent projection, BT reference, and coupled relaxation~\cite{SupplementalMaterial}.

\section*{Data and Code Availability}
The processed data and analysis scripts underlying the reported figures will be made publicly available in Zenodo upon acceptance of the manuscript~\cite{ZenodoDataset}.

\begin{acknowledgments}
This work was supported by the National Research Foundation of Korea(NRF) grant funded by the Korea government(MSIT) (Nos.~RS-2026-25497692 and RS-2025-25418541).
\end{acknowledgments}


\begin{thebibliography}{99}
\color{black}
\hypersetup{urlcolor=black,linkcolor=black,citecolor=black}

\bibitem{Onsager1931I}
\textcolor{black}{L. Onsager. Reciprocal relations in irreversible processes. I. \textit{Phys. Rev.} \textbf{1931}, \textit{37}, 405--426. \url{https://doi.org/10.1103/physrev.37.405}.}


\bibitem{Green1954}
\textcolor{black}{M. S. Green. Markoff random processes and the statistical mechanics of time-dependent phenomena. II. Irreversible processes in fluids. \textit{J. Chem. Phys.} \textbf{1954}, \textit{22}, 398--413. \url{https://doi.org/10.1063/1.1740082}.}


\bibitem{Kubo1957}
\textcolor{black}{R. Kubo. Statistical-mechanical theory of irreversible processes. I. General theory and simple applications to magnetic and conduction problems. \textit{J. Phys. Soc. Jpn.} \textbf{1957}, \textit{12}, 570--586. \url{https://doi.org/10.1143/jpsj.12.570}.}


\bibitem{HansenMcDonald}
J.-P. Hansen; I. R. McDonald.
\textit{Theory of Simple Liquids: With Applications to Soft Matter},
4th ed.; Academic Press: Amsterdam, 2013.


\bibitem{DebyeHuckel1923}
P. Debye; E. H\"uckel. Zur Theorie der Elektrolyte. I.
Gefrierpunktserniedrigung und verwandte Erscheinungen.
\textit{Phys. Z.} \textbf{1923}, \textit{24}, 185--206.


\bibitem{Fong2020Transport}
\textcolor{black}{K. D. Fong; H. K. Bergstrom; B. D. McCloskey; K. K. Mandadapu. Transport phenomena in electrolyte solutions: Nonequilibrium thermodynamics and statistical mechanics. \textit{AIChE J.} \textbf{2020}, \textit{66}, e17091. \url{https://doi.org/10.1002/aic.17091}.}


\bibitem{Roling2024Dynamics}
\textcolor{black}{B. Roling; V. Miß; J. Kettner. Ion dynamics in concentrated electrolyte solutions: Relating equilibrium fluctuations of the ions to transport properties in battery cells. \textit{Energy Environ. Mater.} \textbf{2024}, \textit{7}, e12533. \url{https://doi.org/10.1002/eem2.12533}.}


\bibitem{Balos2020Conductivity}
\textcolor{black}{V. Balos; S. Imoto; R. R. Netz; M. Bonn; D. J. Bonthuis; Y. Nagata; J. Hunger. Macroscopic conductivity of aqueous electrolyte solutions scales with ultrafast microscopic ion motions. \textit{Nat. Commun.} \textbf{2020}, \textit{11}, 1611. \url{https://doi.org/10.1038/s41467-020-15450-2}.}


\bibitem{Caillol1987FiniteWave}
\textcolor{black}{J. M. Caillol. The dielectric constant and the conductivity of an electrolyte solution at finite wave-lengths and frequencies. \textit{Europhys. Lett.} \textbf{1987}, \textit{4}, 159--166. \url{https://doi.org/10.1209/0295-5075/4/2/006}.}


\bibitem{Minh2023ElectricalNoise}
\textcolor{black}{T. H. N. Minh; J. Kim; G. Pireddu; I. Chubak; S. Nair; B. Rotenberg. Electrical noise in electrolytes: A theoretical perspective. \textit{Faraday Discuss.} \textbf{2023}, \textit{246}, 198--224. \url{https://doi.org/10.1039/D3FD00026E}.}


\bibitem{Minh2026Asymmetric}
\textcolor{black}{T. H. N. Minh; S. Varghese; B. Rotenberg. Coupled concentration-charge dynamics in 1:1 electrolytes with unequal diffusion coefficients: Local transient response and fluctuations. \textit{J. Chem. Phys.} \textbf{2026}, \textit{164}, 194109. \url{https://doi.org/10.1063/5.0323816}.}


\bibitem{vanHove1954}
\textcolor{black}{L. Van Hove. Correlations in space and time and Born approximation scattering in systems of interacting particles. \textit{Phys. Rev.} \textbf{1954}, \textit{95}, 249--262. \url{https://doi.org/10.1103/physrev.95.249}.}


\bibitem{deGennes1959}
\textcolor{black}{P. G. De Gennes. Liquid dynamics and inelastic scattering of neutrons. \textit{Physica} \textbf{1959}, \textit{25}, 825--839. \url{https://doi.org/10.1016/0031-8914(59)90006-0}.}


\bibitem{Mori1965}
\textcolor{black}{H. Mori. Transport, collective motion, and Brownian motion. \textit{Prog. Theor. Phys.} \textbf{1965}, \textit{33}, 423--455. \url{https://doi.org/10.1143/ptp.33.423}.}


\bibitem{SvedbergInouye1911}
The Svedberg; K. Inouye. Eine neue Methode zur Prüfung der Gültigkeit des Boyle-Gay-Lussacschen Gesetzes für kolloide Lösungen. \textit{Z. Phys. Chem.} \textbf{1911}, \textit{77U}, 145--191. \href{https://doi.org/10.1515/zpch-1911-7711}{DOI: 10.1515/zpch-1911-7711}.


\bibitem{Smoluchowski1916Counting}
M. von Smoluchowski. Studien über Kolloidstatistik und den Mechanismus der Diffusion. \textit{Kolloid Z.} \textbf{1916}, \textit{18}, 48--54. \url{https://doi.org/10.1007/BF01432660}.


\bibitem{MartinYalcin1980}
P. A. Martin; T. Yalcin. The charge fluctuations in classical Coulomb systems. \textit{J. Stat. Phys.} \textbf{1980}, \textit{22}, 435--463. \url{https://doi.org/10.1007/BF01012866}.


\bibitem{Lebowitz1983Charge}
J. L. Lebowitz. Charge fluctuations in Coulomb systems. \textit{Phys. Rev. A} \textbf{1983}, \textit{27}, 1491--1494. \url{https://doi.org/10.1103/PhysRevA.27.1491}.


\bibitem{KimLuijtenFisher2005}
Y. C. Kim; E. Luijten; M. E. Fisher. Screening in ionic systems: Simulations for the Lebowitz length. \textit{Phys. Rev. Lett.} \textbf{2005}, \textit{95}, 145701. \url{https://doi.org/10.1103/PhysRevLett.95.145701}.


\bibitem{KimFisher2008Charge}
\textcolor{black}{Y. C. Kim; M. E. Fisher. Charge fluctuations and correlation lengths in finite electrolytes. \textit{Phys. Rev. E} \textbf{2008}, \textit{77}, 051502. \url{https://doi.org/10.1103/physreve.77.051502}.}


\bibitem{MinhRotenbergMarbach2023}
\textcolor{black}{T. Hoang Ngoc Minh; B. Rotenberg; S. Marbach. Ionic fluctuations in finite volumes: Fractional noise and hyperuniformity. \textit{Faraday Discuss.} \textbf{2023}, \textit{246}, 225--250. \url{https://doi.org/10.1039/d3fd00031a}.}


\bibitem{Mackay2024}
\textcolor{black}{E. K. R. Mackay; S. Marbach; B. Sprinkle; A. L. Thorneywork. The Countoscope: Measuring self and collective dynamics without trajectories. \textit{Phys. Rev. X} \textbf{2024}, \textit{14}, 041016. \url{https://doi.org/10.1103/physrevx.14.041016}.}


\bibitem{Carter2025}
\textcolor{black}{A. Carter; E. K. R. Mackay; B. Sprinkle; A. L. Thorneywork; S. Marbach. Measuring collective diffusion coefficients by counting particles in boxes. \textit{Soft Matter} \textbf{2025}, \textit{21}, 3991--4002. \url{https://doi.org/10.1039/d4sm01455c}.}


\bibitem{KirkwoodBuff1951}
\textcolor{black}{J. G. Kirkwood; F. P. Buff. The statistical mechanical theory of solutions. I. \textit{J. Chem. Phys.} \textbf{1951}, \textit{19}, 774--777. \url{https://doi.org/10.1063/1.1748352}.}


\bibitem{Kruger2013}
\textcolor{black}{P. Krüger; S. K. Schnell; D. Bedeaux; S. Kjelstrup; T. J. H. Vlugt; J.-M. Simon. Kirkwood–Buff integrals for finite volumes. \textit{J. Phys. Chem. Lett.} \textbf{2013}, \textit{4}, 235--238. \url{https://doi.org/10.1021/jz301992u}.}


\bibitem{Dawass2018}
\textcolor{black}{N. Dawass; P. Krüger; S. K. Schnell; D. Bedeaux; S. Kjelstrup; J. M. Simon; T. J. H. Vlugt. Finite-size effects of Kirkwood–Buff integrals from molecular simulations. \textit{Mol. Simul.} \textbf{2018}, \textit{44}, 599--612. \url{https://doi.org/10.1080/08927022.2017.1416114}.}


\bibitem{SevillaCortes2022}
\textcolor{black}{M. Sevilla; R. Cortes-Huerto. Connecting density fluctuations and Kirkwood–Buff integrals for finite-size systems. \textit{J. Chem. Phys.} \textbf{2022}, \textit{156}, 044502. \url{https://doi.org/10.1063/5.0076744}.}


\bibitem{Simon2022}
\textcolor{black}{J. M. Simon; P. Krüger; S. K. Schnell; T. J. H. Vlugt; S. Kjelstrup; D. Bedeaux. Kirkwood–Buff integrals: From fluctuations in finite volumes to the thermodynamic limit. \textit{J. Chem. Phys.} \textbf{2022}, \textit{157}, 130901. \url{https://doi.org/10.1063/5.0106162}.}


\bibitem{Hermann2026IntensityCountoscope}
\textcolor{black}{S. Hermann; S. S. Banarooei; A. Carter; C. A. Silvera Batista; S. Marbach. The ``Intensity'' Countoscope: Measuring particle dynamics in real space from microscopy images. \textit{arXiv} \textbf{2026}, arXiv:2604.02271. \url{https://doi.org/10.48550/arXiv.2604.02271}.}


\bibitem{Dean1996}
\textcolor{black}{D. S. Dean. Langevin equation for the density of a system of interacting Langevin processes. \textit{J. Phys. A: Math. Gen.} \textbf{1996}, \textit{29}, L613--L617. \url{https://doi.org/10.1088/0305-4470/29/24/001}.}


\bibitem{Illien2025}
\textcolor{black}{P. Illien. The Dean-Kawasaki equation and stochastic density functional theory. \textit{Rep. Prog. Phys.} \textbf{2025}, \textit{88}, 086601. \url{https://doi.org/10.1088/1361-6633/adee2e}.}


\bibitem{DemeryDean2016}
V. Démery; D. S. Dean. The conductivity of strong electrolytes from stochastic density functional theory. \textit{J. Stat. Mech.} \textbf{2016}, 023106. \href{https://doi.org/10.1088/1742-5468/2016/02/023106}{Publisher link}.


\bibitem{BernardJardatRotenbergIllien2023}
\textcolor{black}{O. Bernard; M. Jardat; B. Rotenberg; P. Illien. On analytical theories for conductivity and self-diffusion in concentrated electrolytes. \textit{J. Chem. Phys.} \textbf{2023}, \textit{159}, 164105. \url{https://doi.org/10.1063/5.0165533}.}


\bibitem{Bonneau2024Frequency}
\textcolor{black}{H. Bonneau; Y. Avni; D. Andelman; H. Orland. Frequency-dependent conductivity of concentrated electrolytes: A stochastic density functional theory. \textit{J. Chem. Phys.} \textbf{2024}, \textit{161}, 244501. \url{https://doi.org/10.1063/5.0236073}.}


\bibitem{BonneauDemeryRaphael2025}
H. Bonneau; V. Démery; E. Raphaël. Stationary and transient correlations in driven electrolytes. \textit{J. Stat. Mech.} \textbf{2025}, 033201. \href{https://doi.org/10.1088/1742-5468/adb4ce}{Publisher link}.


\bibitem{IllienCarofRotenberg2024}
\textcolor{black}{P. Illien; A. Carof; B. Rotenberg. Stochastic density functional theory for ions in a polar solvent. \textit{Phys. Rev. Lett.} \textbf{2024}, \textit{133}, 268002. \url{https://doi.org/10.1103/physrevlett.133.268002}.}


\bibitem{VargheseRotenbergIllien2026}
S. Varghese; B. Rotenberg; P. Illien. Solvent-induced memory effects in a
model electrolyte. \textit{arXiv} \textbf{2026}, arXiv:2605.06293v2.
\url{https://doi.org/10.48550/arXiv.2605.06293}.


\bibitem{Kornyshev1983Nonlocal}
\textcolor{black}{A. A. Kornyshev. Non-local dielectric response of a polar solvent and Debye screening in ionic solution. \textit{J. Chem. Soc., Faraday Trans. 2} \textbf{1983}, \textit{79}, 651--661. \url{https://doi.org/10.1039/F29837900651}.}


\bibitem{Berthoumieux2024Nonlinear}
\textcolor{black}{H. Berthoumieux; V. Démery; A. C. Maggs. Nonlinear conductivity of aqueous electrolytes: Beyond the first Wien effect. \textit{J. Chem. Phys.} \textbf{2024}, \textit{161}, 184504. \url{https://doi.org/10.1063/5.0226773}.}


\bibitem{DemeryToquer2026Solvent}
\textcolor{black}{V. Démery; D. Toquer; H. Berthoumieux. Effect of solvent structure on the Wien effect and ionic correlations at the nanoscale. \textit{Faraday Discuss.} \textbf{2026}, \textit{266}, 361--376. \url{https://doi.org/10.1039/d5fd00149h}.}


\bibitem{BhatiaThornton1970}
\textcolor{black}{A. B. Bhatia; D. E. Thornton. Structural aspects of the electrical resistivity of binary alloys. \textit{Phys. Rev. B} \textbf{1970}, \textit{2}, 3004--3012. \url{https://doi.org/10.1103/physrevb.2.3004}.}


\bibitem{CiachGozdzStell2007}
\textcolor{black}{A. Ciach; W. T. Góźdź; G. Stell. Field theory for size- and charge-asymmetric primitive model of ionic systems: Mean-field stability analysis and pretransitional effects. \textit{Phys. Rev. E} \textbf{2007}, \textit{75}, 051505. \url{https://doi.org/10.1103/physreve.75.051505}.}


\bibitem{MarconiTarazona1999}
\textcolor{black}{U. Marini Bettolo Marconi; P. Tarazona. Dynamic density functional theory of fluids. \textit{J. Chem. Phys.} \textbf{1999}, \textit{110}, 8032--8044. \url{https://doi.org/10.1063/1.478705}.}


\bibitem{ArcherEvans2004}
\textcolor{black}{A. J. Archer; R. Evans. Dynamical density functional theory and its application to spinodal decomposition. \textit{J. Chem. Phys.} \textbf{2004}, \textit{121}, 4246--4254. \url{https://doi.org/10.1063/1.1778374}.}


\bibitem{UhlenbeckOrnstein1930}
\textcolor{black}{G. E. Uhlenbeck; L. S. Ornstein. On the theory of the Brownian motion. \textit{Phys. Rev.} \textbf{1930}, \textit{36}, 823--841. \url{https://doi.org/10.1103/physrev.36.823}.}


\bibitem{Straube2020Friction}
\textcolor{black}{A. V. Straube; B. G. Kowalik; R. R. Netz; F. Höfling. Rapid onset of molecular friction in liquids bridging between the atomistic and hydrodynamic pictures. \textit{Commun. Phys.} \textbf{2020}, \textit{3}, 126. \url{https://doi.org/10.1038/s42005-020-0389-0}.}


\bibitem{Ayaz2022Memory}
\textcolor{black}{C. Ayaz; L. Scalfi; B. A. Dalton; R. R. Netz. Generalized Langevin equation with a nonlinear potential of mean force and nonlinear memory friction from a hybrid projection scheme. \textit{Phys. Rev. E} \textbf{2022}, \textit{105}, 054138. \url{https://doi.org/10.1103/physreve.105.054138}.}


\bibitem{CarnahanStarling1969}
\textcolor{black}{N. F. Carnahan; K. E. Starling. Equation of state for nonattracting rigid spheres. \textit{J. Chem. Phys.} \textbf{1969}, \textit{51}, 635--636. \url{https://doi.org/10.1063/1.1672048}.}


\bibitem{SupplementalMaterial}
See Supplemental Material at [URL will be inserted by publisher] for the
simulation and ensemble conventions, static and charge controls, transport
and reciprocal-space tests, and the full ion--solvent decomposition.


\bibitem{joly2006liquid}
\textcolor{black}{L. Joly; C. Ybert; E. Trizac; L. Bocquet. Liquid friction on charged surfaces: From hydrodynamic slippage to electrokinetics. \textit{J. Chem. Phys.} \textbf{2006}, \textit{125}, 204716. \url{https://doi.org/10.1063/1.2397677}.}


\bibitem{Kim2026StructuralDynamicalCrossovers}
\textcolor{black}{D. Kim; T. Kwon; J. Kim. Structural and dynamical crossovers in dense electrolytes. \textit{Phys. Rev. E} \textbf{2026}, \textit{113}, 025405. \url{https://doi.org/10.1103/t1m7-8qm2}.}


\bibitem{WeeksChandlerAndersen1971}
\textcolor{black}{J. D. Weeks; D. Chandler; H. C. Andersen. Role of repulsive forces in determining the equilibrium structure of simple liquids. \textit{J. Chem. Phys.} \textbf{1971}, \textit{54}, 5237--5247. \url{https://doi.org/10.1063/1.1674820}.}


\bibitem{DesernoHolm1998}
\textcolor{black}{M. Deserno; C. Holm. How to mesh up Ewald sums. I. A theoretical and numerical comparison of various particle mesh routines. \textit{J. Chem. Phys.} \textbf{1998}, \textit{109}, 7678--7693. \url{https://doi.org/10.1063/1.477414}.}


\bibitem{Nose1984}
\textcolor{black}{S. Nosé. A unified formulation of the constant temperature molecular dynamics methods. \textit{J. Chem. Phys.} \textbf{1984}, \textit{81}, 511--519. \url{https://doi.org/10.1063/1.447334}.}


\bibitem{Hoover1985}
\textcolor{black}{W. G. Hoover. Canonical dynamics: Equilibrium phase-space distributions. \textit{Phys. Rev. A} \textbf{1985}, \textit{31}, 1695--1697. \url{https://doi.org/10.1103/physreva.31.1695}.}


\bibitem{MartynaTobiasKlein1994}
\textcolor{black}{G. J. Martyna; D. J. Tobias; M. L. Klein. Constant pressure molecular dynamics algorithms. \textit{J. Chem. Phys.} \textbf{1994}, \textit{101}, 4177--4189. \url{https://doi.org/10.1063/1.467468}.}


\bibitem{Cheng2022}
\textcolor{black}{B. Cheng. Computing chemical potentials of solutions from structure factors. \textit{J. Chem. Phys.} \textbf{2022}, \textit{157}, 121101. \url{https://doi.org/10.1063/5.0107059}.}


\bibitem{Helfand1960}
\textcolor{black}{E. Helfand. Transport coefficients from dissipation in a canonical ensemble. \textit{Phys. Rev.} \textbf{1960}, \textit{119}, 1--9. \url{https://doi.org/10.1103/physrev.119.1}.}


\bibitem{McDanielSon2018}
\textcolor{black}{J. G. McDaniel; C. Y. Son. Ion correlation and collective dynamics in BMIM/BF$_4$-based organic electrolytes: From dilute solutions to the ionic liquid limit. \textit{J. Phys. Chem. B} \textbf{2018}, \textit{122}, 7154--7169. \url{https://doi.org/10.1021/acs.jpcb.8b04886}.}


\bibitem{ZenodoDataset}
The Zenodo DOI will be publicly available upon acceptance of the manuscript.


\end{thebibliography}
\end{document}